\documentclass{article}

\PassOptionsToPackage{numbers, compress}{natbib}

\usepackage[preprint]{neurips_2021}

\usepackage[utf8]{inputenc} 
\usepackage[T1]{fontenc}    
\usepackage{hyperref}       
\usepackage{url}            
\usepackage{booktabs}       
\usepackage{amsfonts}       
\usepackage{nicefrac}       
\usepackage{graphicx} \graphicspath{ {figures/} }
\usepackage{microtype}      
\usepackage{xcolor}         
\usepackage{amsmath}
\usepackage{multirow}
\usepackage{multicol}
\usepackage{subcaption, multicol, multirow}

\newcommand{\repeatthanks}{\textsuperscript{\thefootnote}}
\DeclareMathOperator*{\argmax}{arg\,max}

\title{3N-GAN: Semi-Supervised Classification of X-Ray Images with a 3-Player Adversarial Framework}

\author{%
  Shafin Haque\thanks{Equal Contribution} \\
  Saratoga High School\\
  Saratoga, CA 95070 \\
  \texttt{shafin1025@gmail.com} \\
   \And
  Ayaan Haque\repeatthanks \\
  Saratoga High School\\
  Saratoga, CA 95070 \\
  \texttt{ayaanzhaque@gmail.com} \\
}

\begin{document}

\maketitle

\begin{abstract}
  The success of deep learning for medical imaging tasks, such as classification, is heavily reliant on the availability of large-scale datasets. However, acquiring datasets with large quantities of labeled data is challenging, as labeling is expensive and time-consuming. Semi-supervised learning (SSL) is a growing alternative to fully-supervised learning, but requires unlabeled samples for training. In medical imaging, many datasets lack unlabeled data entirely, so SSL can't be conventionally utilized. We propose 3N-GAN, or 3 Network Generative Adversarial Networks, to perform semi-supervised classification of medical images in fully-supervised settings. We incorporate a classifier into the adversarial relationship such that the generator trains adversarially against \textit{both} the classifier and discriminator. Our preliminary results show improved classification performance and GAN generations over various algorithms. Our work can seamlessly integrate with numerous other medical imaging model architectures and SSL methods for greater performance. \footnote{Our code, containing licenses, hyperparameters, and data, will be made available after review.}
\end{abstract}

\section{Introduction}

Medical image analysis via deep learning is heavily reliant on large-scale, labeled datasets \cite{shen2017deep}. Semi-supervised learning (SSL) has gained attention as an alternative to fully-supervised learning. SSL is for tasks which have datasets containing unlabeled samples, and there have been many works in such direction \cite{anwar2018medical, Liu2020May, haque2021multimix, gu2019conext, liu2020semi}. However, in medical imaging, datasets which are fully labeled but extremely low in samples are more common. For example, x-rays are most often taken for diagnostic purposes, so they are generally accompanied with labels. The difficulty lies in aggregating these x-rays.

Restricted, fully-supervised datasets can be aided by deep generative models, as they can generate artificial samples to supplement full-supervised training. Generative Adversarial Networks \cite{Goodfellow2014Jun} are a prominent form of generative models where a discriminator and generator work adversarially towards one another. The generator is trained to produce realistic images by learning to replicate the data distribution while the discriminator is trained to distinguish between real and fake samples. Structured as a 2-player game, only very few papers construct an adversarial framework with an additional neural classifier \cite{li2017triple, shen2018faceid}. To our best knowledge, none utilize classifier predictions on unlabeled generations to update the generator, especially using pseudo-labels \cite{lee2013pseudo}. Also to our best knowledge, no methods in medical imaging use GANs to generate artificial samples for semi-supervised classification.

We introduce 3N-GAN, or 3 Network Generative Adversarial Network, for semi-supervised classification of X-ray images using generated samples as supplemental data. Our results confirm that our 3-player adversarial framework with innovative adversarial loss functions improves classifier and generator performance over baseline models at varying levels of supervision.

\section{Methods}

\label{sec:method}
\begin{figure}
    \centering
    \includegraphics[width=0.8\linewidth]{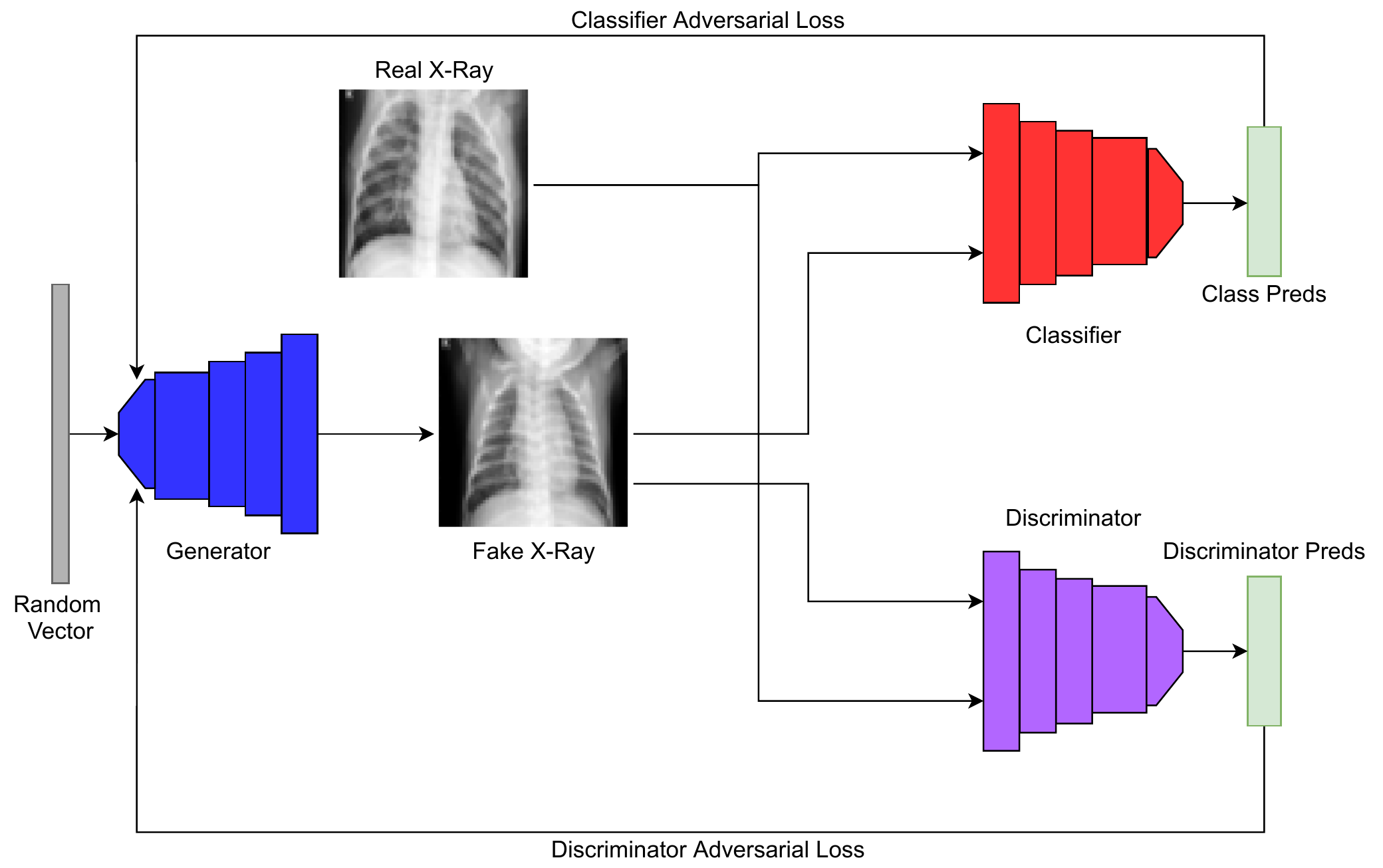}
    \caption{Schematic of 3N-GAN. The generator (blue) produces fake X-Ray images, which are unlabeled. The classifier (red) is trained on both real and fake images and uses pseudo-labeling for semi-supervised classification. The generator is updated on the classifier adversarial loss (based on the classifier-produced pseudo-label) and the discriminator adversarial loss.}
    \label{fig:model}
\end{figure}

Figure \ref{fig:model} displays the adversarial training procedure for 3N-GAN. This method can be viewed as an extension of \cite{Haque_2021}, which is only published as a 2-page abstract. Our method, 3N-GAN, introduces multiple significant improvements, such as truly incorporating the classifier into the adversarial framework, which the previous method did not attempt.

All 3 networks are trained simultaneously. The discriminator is trained conventionally. The generator is given a random latent vector as an input and outputs fake images. Contrary to many semi-supervised GAN classification methods \cite{improvedgans}, we separate the discriminator and classifier. Achieving two tasks with a single network when the tasks are not very related, such as discrimination and classification, may not be optimal as the network must approximate two distributions \cite{Haque_2021}.

The classifier is simultaneously trained on real images and artificial images produced by the generator. Generated images act as supplemental data for a classifier, increasing the amount of data available to the classifier. GAN generated samples do not have labels, requiring the use of semi-supervised algorithms. For semi-supervised classification, we use both pseudo-labeling \cite{lee2013pseudo} and KL divergence loss, which has not been used with GAN-based semi-supervised classification before. KL divergence loss ensures consistency by penalizing divergence between predictions on real and generated samples.

The classification loss objective

\begin{equation}
    L_c(y, \hat{y}, \hat{y_g}) = L_s(\hat{y}, y) + \lambda L_u(\hat{y_g}, \argmax(\hat{y_g}) > \tau) + \alpha L_{kl}(\hat{y}, \hat{y_g})
\end{equation}

has a supervised component $L_c$ for the labels ($y$) and predictions ($\hat{y}$) of real samples, an unsupervised component $L_u$ for the predictions on generated samples ($\hat{y_g}$) and their pseudo-labels ($\argmax(\hat{y_g}) > \tau$), and a second unsupervised component computing KL divergence loss $L_{kl}$ between the predictions on real samples and GAN generated samples. $\lambda$ and $\alpha$ are unsupervised loss weights and $\tau$ is the pseudo-labeling threshold.

To incorporate the classifier into the adversarial framework, we update the generator from the unsupervised classification loss. The generator now has a discriminator adversarial loss and a classifier adversarial loss. The generator objective 

\begin{equation}
    L_g(\hat{y_d}, \hat{y_g}) = L_d(\hat{y_d}, 1) + \lambda L_{cu}(\hat{y_g}, \argmax(\hat{y_g}) > \tau)
\end{equation}

has the discriminator adversarial loss $L_d$ based on the discriminator predictions ($\hat{y_d}$) on generated images and the classifier adversarial loss $L_{cu}$ which is identical to the unsupervised classification loss from $L_c$. 

The classifier adversarial loss trains the generator to generate images whose class can be discerned accurately. Certain features that distinguish between classes, such as lung inflammations, may be more accurately produced in the generator's samples. The classifier and generator provide feedback to one another, completing the 3-player adversarial framework. To our best knowledge, updating the generator on classification predictions using a pseudo-label is entirely novel and unavailable in literature.

\section{Results and Conclusion}

We experiment with a binary pneumonia classification dataset (CheX) \cite{kermany2018identifying}. The dataset contains a total of 5,863 X-Rays with 624 as an external validation set. We perform experiments at various quantities of training data: 200, 500, 750, 1000, and 2000 X-rays (even split between classes). We use the DCGAN \cite{radford2015unsupervised} implementation for our generator and discriminator. Our classifier architecture is the DC discriminator. All inputs were normalized, gray-scaled, and resized to 64 $\times$ 64 $\times$ 1 before training. Each experiment was trained for 100 epochs with mini-batch size 10 and repeated 5 times. We compare against: a vanilla classifier, a multi-tasking discriminator, and EC-GAN. We set $\tau$ = 0.9, $\alpha$, = 0.3 and $\lambda$ = 0.01 through tuning experiments. 

\begin{table}
    \setlength{\tabcolsep}{4pt}
    \centering
    \caption{Accuracy metrics of baselines against 3N-GAN at various quantities of data. The best scores at each dataset size are bolded.}
    \resizebox{0.6\linewidth}{!}{
    \begin{tabular}{@{} cc c cccc@{}}
    \toprule
    \multirow{2}{*}{Model} 
    &&
    \multicolumn{5}{c}{Dataset Size}
    \\
    \cmidrule{3-7}
    && 200 & 500 & 750 & 1000 & 2000 \\
    \midrule
    Vanilla Classifier && 85.04 & 87.82 & 89.17 & 90.8 & 91.67 \\
    Multi-Tasking Discriminator \cite{improvedgans} &&  89.06 & 92.12 & 91.85 & 94.12 & 94.85 \\
    EC-GAN \cite{Haque_2021} &&  90.3 & 92.44 & 93.10 & 94.95 & 95.35 \\
    3N-GAN &&  \textbf{91.52} & \textbf{93.49} & \textbf{94.16} & \textbf{94.88} & \textbf{96.03} \\
    \bottomrule
    \end{tabular}
    }
    \label{tab:final-scores}
\end{table}

\begin{figure}
    \centering
    \resizebox{0.75\linewidth}{!}{
    \subcaptionbox{DCGAN Generated X-Rays}{\includegraphics[width=0.32\linewidth, trim={100 30 100 30}, clip]{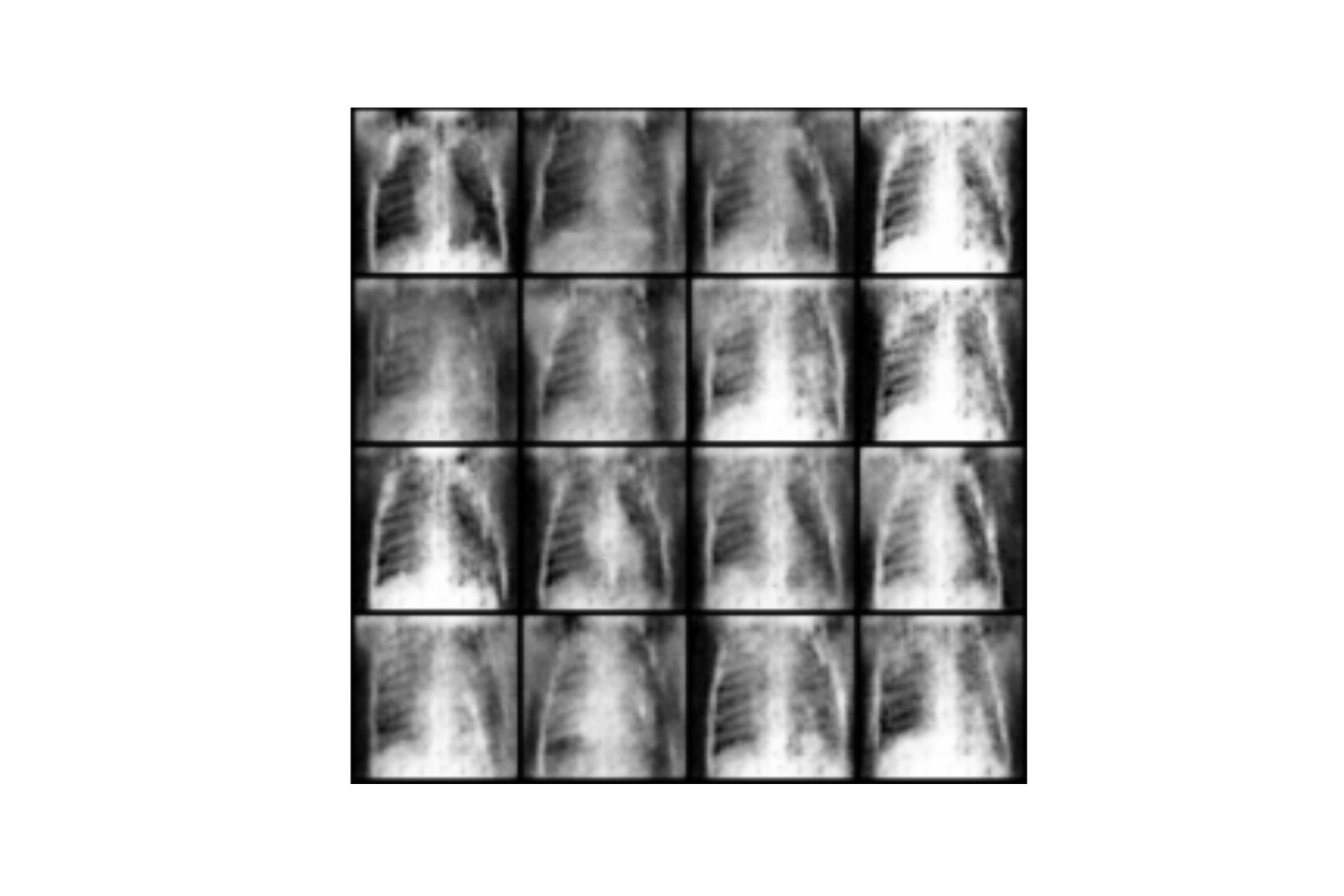}}
    \subcaptionbox{3N-GAN Generated X-Rays}{\includegraphics[width=0.32\linewidth, trim={100 30 100 30}, clip]{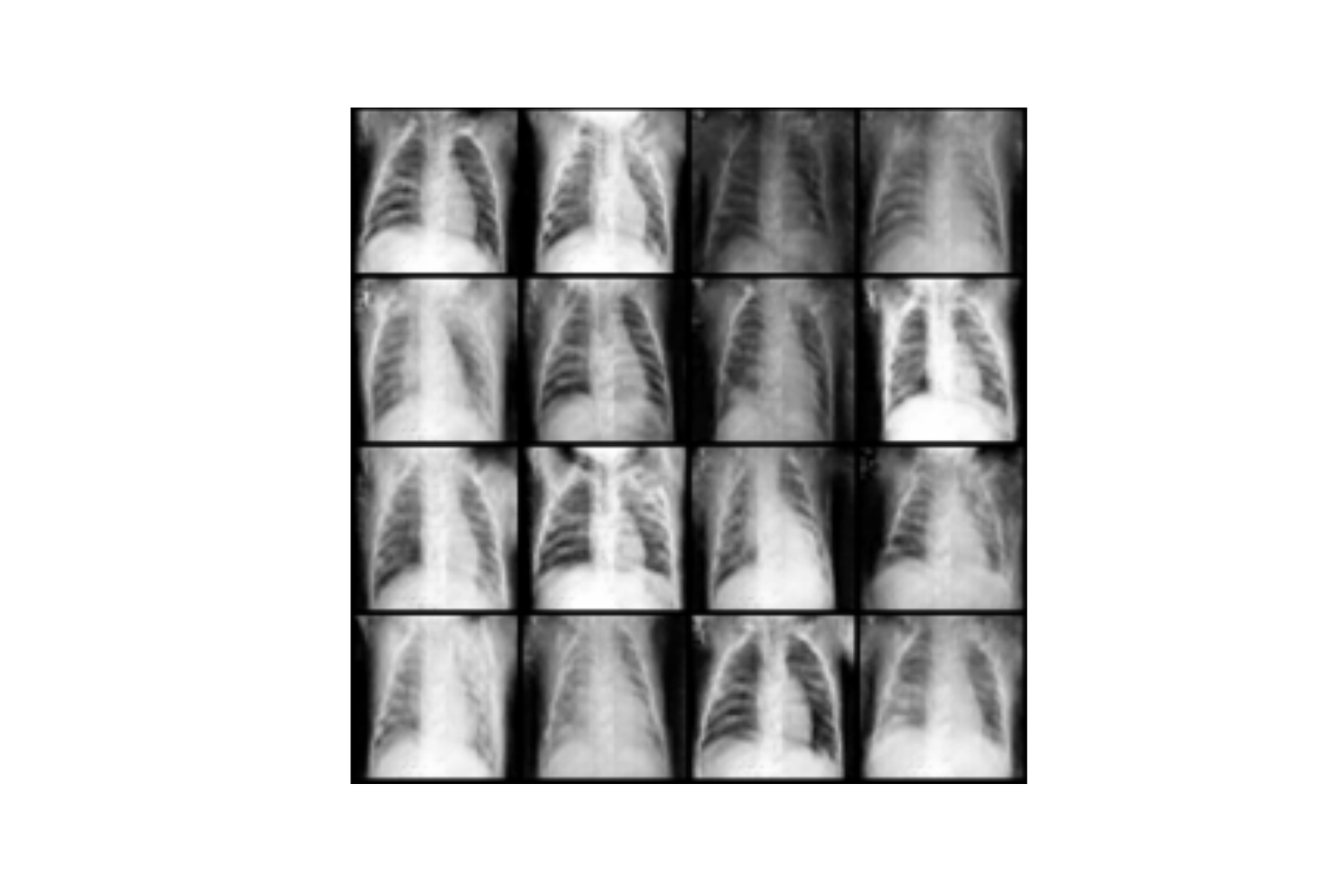}}
    \subcaptionbox{Real X-Rays}{\includegraphics[width=0.32\linewidth, trim={100 30 100 30}, clip]{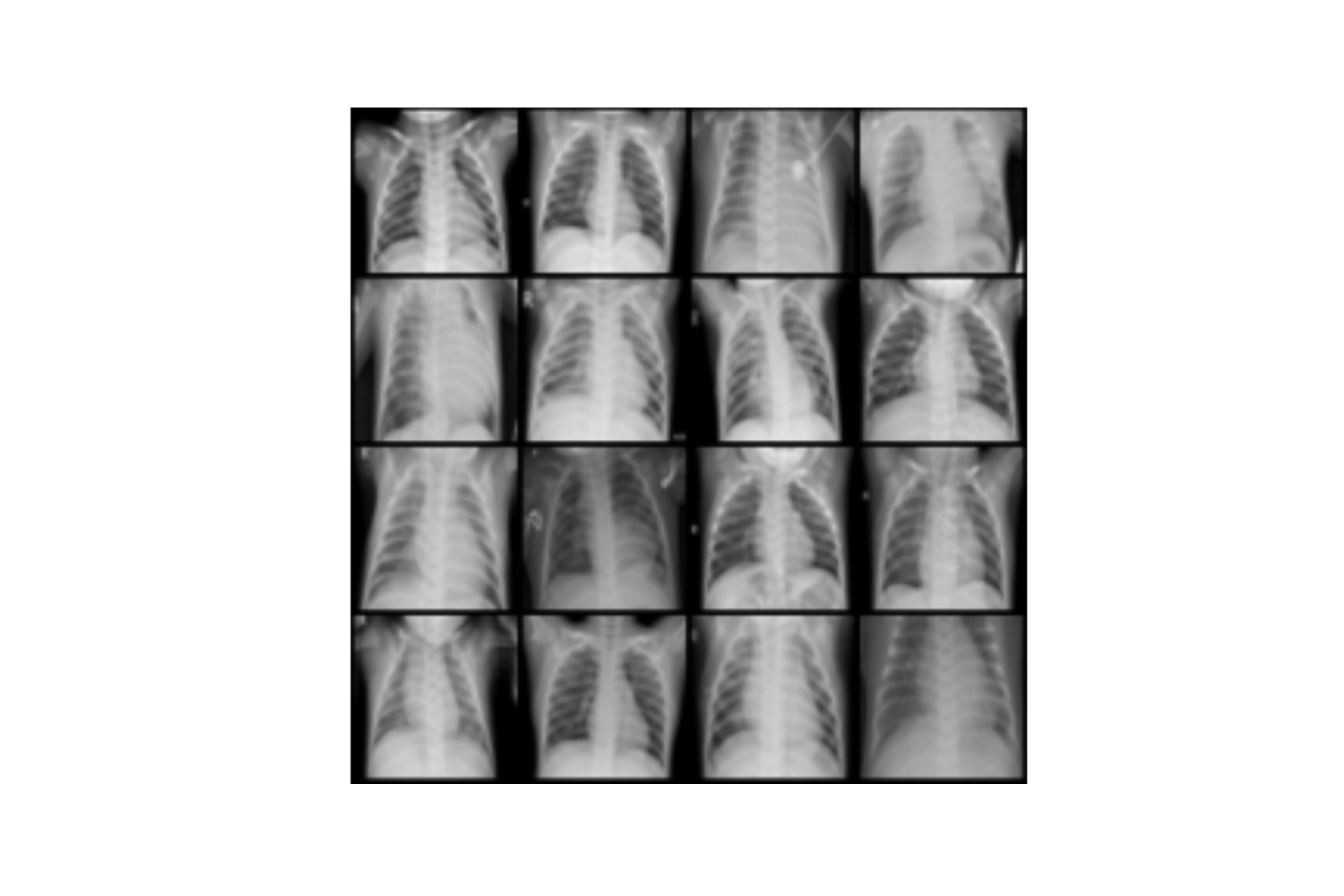}}
    }
    \caption{Generated X-rays from DCGAN and our model, 3N-GAN (using the same input vector) compared to real X-Rays.}
    \label{fig:gan-images}
\end{figure}

Table \ref{tab:final-scores} displays the accuracy scores for diagnosing pneumonia of each model at the 5 different data settings. The results confirm the superiority of 3N-GAN over the other methods, as our model trained with just 200 images almost reaches the performance of a vanilla classifier trained with 2000 images. Compared to EC-GAN, we have slightly increased accuracy ($\approx$ 1\% raw score), proving the effectiveness of the adversarial classifier loss. Figure \ref{fig:gan-images} compares fake X-rays from DCGAN and 3N-GAN to real X-rays. Visually, 3N-GAN generations are less blurry and more defined, proving the importance of our contributions. Moreover, key structures such as ribs and organs are sharp and present in 3N-GAN generations, and they resemble real X-rays more accurately than a DCGAN. 

Ultimately, our preliminary results are quite promising, as 3N-GAN outperforms related methods in both diagnostic classification and x-ray image generation. In 3N-GAN, our classifier performance increases because the supplemental images supplied by the generator are of higher, improved quality. Our future work will include evaluations against even more methods and investigate additional changes to improve image generation.

\section{Potential Negative Impacts}
Potential negative impacts of our work include deepfake ethical violations related to GANs. GANs have been used for deepfakes \cite{korshunov2018deepfakes}, which are AI generated media to replicate and falsify a particular individual's likeness. Our GAN could be exploited to generate fake X-ray scans for patients. Synthetic X-rays could be used for false diagnostic results by patients or medical professionals. Fairness issues would also arise if our work were to be used in a clinical setting, such as determining which patients receive priority access to deep learning-based diagnostics. Regarding privacy, collecting patient data for deep learning-based methods poses patient privacy concerns. Another major concern results from false-positive or false-negative diagnoses on X-ray scans. While recent advancements suggest deep learning-based medical imaging analysis is reaching performance levels on par with professionals, if our classification algorithm predicted incorrect diagnoses, major liability and ethical issues would arise. This is a common challenge for deep learning medical imaging diagnostic models \cite{aggarwal2021diagnostic}. By our interpretation, our work does not introduce any unique negative impacts aside from the long-existing challenges of deep learning-based medical imaging analysis. However, these challenges must be addressed before algorithms such as ours could be applied in clinical practice.

\bibliographystyle{plainnat}
\bibliography{refs}

\end{document}